\documentstyle[12pt]{article}
\makeatletter

\@addtoreset{equation}{section}
\makeatother

\newcommand{\beq}{\begin{equation}}
\newcommand{\eeq}{\end{equation}}
\newcommand{\beqa}{\begin{eqnarray}}
\newcommand{\eeqa}{\end{eqnarray}}

\begin{document}

\begin{center}

{\Large\bf On the Finiteness of the ${\cal N}=4$ SUSY Nonlinear Sigma Model
in Three Dimensions}

\vspace{2cm}
{\large
Takeo Inami,
Yorinori Saito
and
Masayoshi Yamamoto
}

\vspace{1cm}
{\it Department of Physics,
Faculty of Science and Engineering,
Chuo University\\
Kasuga, Bunkyo-ku, Tokyo 112-8551, Japan}

\end{center}

\vspace{2cm}
\begin{abstract}
We construct the ${\cal N}=4$ supersymmetric nonlinear sigma model in three dimensions 
which can be expanded in $1/N$.
We evaluate the effective action at leading order in the $1/N$ expansion
and show the finiteness of the model to this order.
\end{abstract}

\newpage

\section{Introduction}

Supersymmetric field theories have softer ultraviolet behavior than
non-supersymmetric theories.
The structure of ultraviolet divergence cancellations
in supersymmetric field theories including non-renormalizable theories
is reviewed in Ref. \cite{Howe}.
In renormalizable theories,
the ${\cal N}=4$ supersymmetric Yang-Mills theory in four dimensions
and the ${\cal N}=4$ supersymmetric nonlinear sigma model in two dimensions
were found to be finite to all orders in perturbation theory.
In non-renormalizable theories,
the maximal super Yang-Mills theories in $d=5$, $6$, and $7$
were found to be divergent
at four, three, and two loops respectively.

Three-dimensional nonlinear sigma models are perturbatively non-renormalizable,
but they are argued to be renormalizable
in the $1/N$ expansion \cite{Arefeva1,Arefeva2,Vasilev,Rosenstein}.
The $O(N)$ and $CP^{N-1}$ sigma models were studied to next-to-leading order in $1/N$
and their $\beta$-functions were determined to this order \cite{Rosenstein,Cant}.
Supersymmetry appears to play a role in controlling the $\beta$-function
through UV divergences in this class of field theories.
In the ${\cal N}=1$ supersymmetric $O(N)$ nonlinear sigma model,
the next-to-leading order term in the $\beta$-function
(the part due to logarithmic divergences)
turned out to be absent \cite{Koures}.
In the ${\cal N}=2$ supersymmetric $CP^{N-1}$ sigma model,
the next-to-leading order term in the $\beta$-function
(due to both logarithmic and power divergences)
was found to vanish \cite{Ciuchini,Inami}.
These results are reminiscent of the UV properties of
${\cal N}=1$ and ${\cal N}=2$ supersymmetric gauge theories
in four dimensions.

It is very interesting to study whether the three-dimensional nonlinear sigma model with higher
(i.e., ${\cal N}=4$)
extended supersymmetry has better UV properties.
In this view we study the UV properties of the ${\cal N}=4$ supersymmetric nonlinear sigma model
in three dimensions using the $1/N$ expansion.
One model of this kind which comes to our mind is the supersymmetric $HP^{N-1}$ model
(an extension of $O(N)$ and $CP^{N-1}$ models),
but this model is known to be a consistent model
only when it is coupled to supergravity \cite{Galicki}.
In four dimensions we have the ${\cal N}=2$ supersymmetric nonlinear sigma model
employing the cotangent bundle of $CP^{N-1}$ \cite{Curtright,Rocek}.
This model can be used to construct the ${\cal N}=4$ supersymmetric nonlinear sigma model
in three dimensions
which serves the purpose of our $1/N$ study.
In this letter we evaluate the effective action at leading order in $1/N$
and find that the model is finite to this order.

\section{The ${\cal N}=4$ SUSY Nonlinear Sigma Model in Three Dimensions}

We start by considering the ${\cal N}=2$ supersymmetric nonlinear sigma model in four dimensions.
This model has been constructed both in ${\cal N}=1$ superfields \cite{Rocek}
and in the component language \cite{Curtright}.
We follow the work of Curtright and Freedman \cite{Curtright}.
The model has $2N$ complex scalar fields $\phi^\alpha_i$ and $N$ Dirac fields $\psi^\alpha$,
the superpartnars of $\phi^\alpha_i$,
where $i=1,2$ and $\alpha=1,\dots,N$.
These fields are subject to the constraints
\beqa
&&\bar{\phi}^\alpha_i (\sigma_I)_{ij}\phi^\alpha_j=b_I,
~~~I=1, 2, 3,
\label{4dconstraint1}\\
&&R(\phi^\alpha_i\psi^{C\alpha}-\epsilon_{ij}\bar{\phi}^\alpha_j\psi^\alpha)
+L(\bar{\phi}^\alpha_i\psi^\alpha-\epsilon_{ij}\phi^\alpha_j\psi^{C\alpha})=0,
\label{4dconstraint2}
\eeqa
where $\sigma_I$ are the Pauli matrices and $b_I$ is a fixed constant vector. 
$R,L=\frac{1}{2}(1\pm \Gamma_5)$ are the projection operators
and $\epsilon_{ij}=\frac{1}{2}[(-)^j-(-)^i]$.
In this letter we use $\Gamma$'s for the Dirac matrices in four dimensions.
The action is given by
\beq
S=\int d^4x [\overline{D_\mu\phi^\alpha_i}D^\mu\phi^\alpha_i
+i\bar{\psi}^\alpha\Gamma_\mu D^\mu\psi^\alpha
+\rho\bar{\psi}^\alpha\psi^\alpha+i\sigma\bar{\psi}^\alpha\Gamma_5\psi^\alpha
-(\rho^2+\sigma^2)\bar{\phi}^\alpha_i\phi^\alpha_i],
\label{4daction}
\eeq
where $D_\mu=\partial_\mu+iA_\mu$,
and $\rho$, $\sigma$, and $A_\mu$ are real auxiliary fields.
The target space of the model is known to be a Calabi manifold with hyper-K\"ahler metric
on the cotangent bundle of $CP^{N-1}$ \cite{Galicki,Alvarez}.

The action and the constraints are invariant under the pair of supersymmetry transformations
\beqa
&&\delta\phi^\alpha_i=\bar{\epsilon}_i L\psi^\alpha
-\epsilon_{ij}\bar{\epsilon}_j R\psi^\alpha,
\label{4dsusyb}\\
&&\delta\psi^\alpha=(-i\Gamma_\mu D^\mu+\rho-i\Gamma_5\sigma)
(R\phi^\alpha_i\epsilon_i-L\epsilon_{ij}\phi^\alpha_i\epsilon_j),
\label{4dsusyf}
\eeqa
provided we use the on-shell values for the auxiliary fields $\rho$, $\sigma$ and $A_\mu$.
Here $\epsilon_i$ are Majorana spinor parameters.
The supersymmetry transformations of the auxiliary fields need not be specified
in the present on-shell formulation.

We construct the ${\cal N}=4$ supersymmetric nonlinear sigma model in three dimensions
by dimensional reduction of the four-dimensional ${\cal N}=2$ model considered above.
We express the four-dimensional Dirac matrices as
\beqa
\Gamma_\mu=\gamma_\mu\otimes\sigma_2,~~~\mu=0,1,2,
~~~\Gamma_3=i\otimes\sigma_3,~~~\Gamma_5=-1\otimes\sigma_1,
\label{gamma}
\eeqa
where $\gamma_\mu$ are the Dirac matrices in three dimensions
and they are given by $\gamma_0=\sigma_2$, $\gamma_1=i\sigma_3$ and $\gamma_2=i\sigma_1$.
We write the four-component Dirac spinors in four dimensions as
\beq
\psi^\alpha=\left(
\begin{array}{c}
\psi^\alpha_1\\
i\psi^\alpha_2
\end{array}
\right),
~~~\psi^{C\alpha}=\left(
\begin{array}{c}
\psi^{*\alpha}_1\\
i\psi^{*\alpha}_2
\end{array}
\right),
\label{spinor}
\eeq
where $\psi^\alpha_1$ and $\psi^\alpha_2$ are two-component complex spinors in three dimensions.
We substitute (\ref{gamma}) and (\ref{spinor})
into (\ref{4dconstraint2}) and (\ref{4daction})
and pick up only the zero mode (with respect to the third space coodinate) parts of the fields.
We obtain the three-dimensional action
\beqa
S&=&\int d^3x [\overline{D_\mu\phi^\alpha_i}D^\mu\phi^\alpha_i
+i\bar{\psi}^\alpha_i\gamma_\mu D^\mu\psi^\alpha_i
+\tau\bar{\psi}^\alpha_1\psi^\alpha_2+\bar{\tau}\bar{\psi}^\alpha_2\psi^\alpha_1
-\sigma(\bar{\psi}^\alpha_1\psi^\alpha_1-\bar{\psi}^\alpha_2 \psi^\alpha_2)
\nonumber\\
&&-(\bar{\tau}\tau+\sigma^2)\bar{\phi}^\alpha_i\phi^\alpha_i],
\label{3daction}
\eeqa
where $\tau=\rho-iA_3$ is a complex scalar field.
As for the constraints,
the direction of $b_I$ is immaterial,
and we choose the bosonic constraint (\ref{4dconstraint1}) to be $b_I=(0,0,N/g)$,
where $g$ is the coupling constant of the model.
The costraints now read  
\beqa
&&\bar{\phi}^\alpha_1 \phi^\alpha_1-\bar{\phi}^\alpha_2\phi^\alpha_2=N/g,
~~~\bar{\phi}^\alpha_1\phi^\alpha_2=0,
\label{3dconstraint1}\\
&&\bar{\phi}^\alpha_1\psi^\alpha_1-i\phi^\alpha_2\psi^{*\alpha}_2=0,
~~~\bar{\phi}^\alpha_1\psi^\alpha_2+i\phi^\alpha_2\psi^{*\alpha}_1=0.
\label{3dconstraint2}
\eeqa

The supersymmetry transformations in three dimensions can be obaind
by dimensional reduction of the supersymmetry transformations (\ref{4dsusyb}) and (\ref{4dsusyf}).
To this end we decompose the two four-component Majorana spinor parameters $\epsilon_i$
into four two-component Majorana spinor parameters $\epsilon^1_i$ and $\epsilon^2_i$
in the same way as (\ref{spinor}).
The three-dimensional supersymmetry transformations are then given by
\beqa
&&\delta\phi^\alpha_i=-\frac{1}{2}i[\bar{\zeta}^*_i (\psi^\alpha_1+i\psi^\alpha_2)
+\epsilon_{ij}\bar{\zeta}_j (\psi^\alpha_1-i\psi^\alpha_2)],
\label{3dsusyb}\\
&&\delta\psi^\alpha_1=\frac{1}{2}[(\gamma_\mu D^\mu-i\sigma+\tau)\phi^\alpha_i\zeta^*_i
+(\gamma_\mu D^\mu-i\sigma-\tau)\epsilon_{ij}\phi^\alpha_i\zeta_j],
\label{3dsusyf1}\\
&&\delta\psi^\alpha_2=-\frac{1}{2}i[(\gamma_\mu D^\mu+i\sigma-\bar{\tau})\phi^\alpha_i\zeta^*_i
-(\gamma_\mu D^\mu+i\sigma+\bar{\tau})\epsilon_{ij}\phi^\alpha_i\zeta_j],
\label{3dsusyf2}
\eeqa
where $\zeta_i=\epsilon^1_i+i\epsilon^2_i$ are complex spinor parameters.
The supersymmetry transformations of the auxiliary fields $\tau$, $\sigma$ and $A_\mu$
need not be specified for the same reason as given previously.

We have derived the three-dimensional ${\cal N}=4$ supersymmetric nonlinear sigma model (\ref{3daction})
with (\ref{3dconstraint1}) and (\ref{3dconstraint2}).
Introducing the Lagrange multiplier fields $\alpha$, $\beta$, $c$ and $e$,
the Euclidean action is written as
\beqa
S&=&\int d^3x [\overline{D_\mu\phi^\alpha_i}D_\mu\phi^\alpha_i
+i\bar{\psi}^\alpha_i\gamma_\mu D_\mu\psi^\alpha_i
-\tau\bar{\psi}^\alpha_1\psi^\alpha_2-\bar{\tau}\bar{\psi}^\alpha_2\psi^\alpha_1
+\sigma(\bar{\psi}^\alpha_1\psi^\alpha_1-\bar{\psi}^\alpha_2\psi^\alpha_2)
\nonumber\\
&&+(\bar{\tau}\tau+\sigma^2)\bar{\phi}^\alpha_i\phi^\alpha_i
-\alpha(\bar{\phi}^\alpha_1\phi^\alpha_1-\bar{\phi}^\alpha_2\phi^\alpha_2-N/g)
-\beta\bar{\phi}^\alpha_1\phi^\alpha_2-\bar{\beta}\bar{\phi}^\alpha_2\phi^\alpha_1
+\bar{\phi}^\alpha_1\bar{c}\psi^\alpha_1+\phi^\alpha_1\bar{\psi}^\alpha_1 c
\nonumber\\
&&+i\bar{\phi}^\alpha_2\bar{c}^*\psi^\alpha_2-i\phi^\alpha_2\bar{\psi}^\alpha_2 c^*
+\bar{\phi}^\alpha_1\bar{e}\psi^\alpha_2+\phi^\alpha_1\bar{\psi}^\alpha_2 e
-i\bar{\phi}^\alpha_2\bar{e}^*\psi^\alpha_1+i\phi^\alpha_2\bar{\psi}^\alpha_1 e^*].
\label{fullaction}
\eeqa
The model contains seven kinds of auxiliary fields:
a $U(1)$ vector $A_\mu$, two complex scalars $\tau$, $\beta$,
two real scalars $\sigma$, $\alpha$,
and two complex spinors $c$, $e$.

\section{The Leading Order}

Integrating over the fields $\phi^\alpha_i$, $\bar{\phi}^\alpha_i$,
$\psi^\alpha_i$ and $\bar{\psi}^\alpha_i$,
we obtain from (\ref{fullaction}) the effective action
\beq
S_{eff}=N{\rm Tr}\ln\Delta_{B1}+N{\rm Tr}\ln\Delta_{B2}
-N{\rm Tr}\ln\Delta_{F1}-N{\rm Tr}\ln\Delta_{F2}
+\frac{N}{g}\int d^3x ~\alpha+\dots,
\label{effaction}
\eeq
where
\beqa
\Delta_{F1}&=&i\gamma_\mu D_\mu+\sigma,
\label{deltaf1}\\
\Delta_{F2}&=&i\gamma_\mu D_\mu-\sigma-\bar{\tau}\Delta_{F1}^{-1}\tau,
\label{deltaf2}\\
\Delta_{B1}&=&-D_\mu D_\mu+\bar{\tau}\tau+\sigma^2-\alpha
-\bar{c}\Delta_{F1}^{-1}c
-(\bar{e}+\bar{c}\Delta_{F1}^{-1}\tau)\Delta_{F2}^{-1}
(e+\bar{\tau}\Delta_{F1}^{-1}c)
\label{deltab1}\\
\Delta_{B2}&=&-D_\mu D_\mu+\bar{\tau}\tau+\sigma^2+\alpha
-\bar{e}^* \Delta_{F1}^{-1}e^*
-(\bar{c}^*-\bar{e}^* \Delta_{F1}^{-1}\tau)\Delta_{F2}^{-1}
(c^*-\bar{\tau}\Delta_{F1}^{-1}e^*)
\nonumber\\
&&-[\bar{\beta}-i\bar{e}^* \Delta_{F1}^{-1}c 
+i(\bar{c}^*-\bar{e}^*\Delta_{F1}^{-1}\tau)\Delta_{F2}^{-1}
(e+\bar{\tau}\Delta_{F1}^{-1}c)]\Delta_{B1}^{-1}
[\beta+i\bar{c}\Delta_{F1}^{-1}e^*
\nonumber\\
&&-i(\bar{e}+\bar{c}\Delta_{F1}^{-1}\tau)\Delta_{F2}^{-1}
(c^*-\bar{\tau}\Delta_{F1}^{-1}e^*)].
\label{deltab2}
\eeqa
Setting all fields to constants,
we obtain the effective potential
\beqa
\frac{V}{N}&=&\int\frac{d^3k}{(2\pi)^3}
\biggl[\ln (k^2+\langle\bar{\tau}\rangle\langle\tau\rangle+\langle\sigma\rangle^2-\langle\alpha\rangle)
\nonumber\\
&&+\ln \biggl(k^2+\langle\bar{\tau}\rangle\langle\tau\rangle+\langle\sigma\rangle^2+\langle\alpha\rangle
+\frac{\langle\bar{\beta}\rangle\langle\beta\rangle}
{k^2+\langle\bar{\tau}\rangle\langle\tau\rangle+\langle\sigma\rangle^2-\langle\alpha\rangle}\biggr)
\nonumber\\
&&-{\rm tr}\ln (-k\!\!\!/+\langle\sigma\rangle)
-{\rm tr}\ln \biggl(-k\!\!\!/-\langle\sigma\rangle-\frac{\langle\bar{\tau}\rangle\langle\tau\rangle}
{-k\!\!\!/+\langle\sigma\rangle}\biggr)\biggr]
\nonumber\\
&&+(\langle\bar{\tau}\rangle\langle\tau\rangle+\langle\sigma\rangle^2)
(\bar{v}_1 v_1+\bar{v}_2  v_2)
-\langle\alpha\rangle (\bar{v}_1 v_1-\bar{v}_2 v_2-1/g)
\nonumber\\
&&-\langle\beta\rangle\bar{v}_1 v_2-\langle\bar{\beta}\rangle\bar{v}_2 v_1
\nonumber\\
&=&\int\frac{d^3k}{(2\pi)^3}
[\ln ((k^2+\langle\bar{\tau}\rangle\langle\tau\rangle+\langle\sigma\rangle^2)^2
-\langle\alpha\rangle^2+\langle\bar{\beta}\rangle\langle\beta\rangle)
-\ln (k^2+\langle\bar{\tau}\rangle\langle\tau\rangle+\langle\sigma\rangle^2)^2]
\nonumber\\
&&+(\langle\bar{\tau}\rangle\langle\tau\rangle+\langle\sigma\rangle^2)
(\bar{v}_1 v_1+\bar{v}_2  v_2)
-\langle\alpha\rangle (\bar{v}_1 v_1-\bar{v}_2 v_2-1/g)
\nonumber\\
&&-\langle\beta\rangle\bar{v}_1 v_2-\langle\bar{\beta}\rangle\bar{v}_2 v_1, 
\label{potential}
\eeqa
where $v_i=\langle\phi^N_i\rangle/\sqrt{N}$.
The vacuum expectation values of fields which are not in (\ref{potential}) have been set to zero.

The vacuum of the model is fixed by the saddle point conditions
\beqa
&&\frac{1}{N}\frac{\partial V}{\partial\langle\tau\rangle}
=\langle\bar{\tau}\rangle(\bar{v}_1 v_1+\bar{v}_2 v_2)=0,
\label{saddle1}\\
&&\frac{1}{N}\frac{\partial V}{\partial\langle\sigma\rangle}
=2\langle\sigma\rangle(\bar{v}_1 v_1+\bar{v}_2 v_2)=0,
\label{saddle2}\\
&&\frac{1}{N}\frac{\partial V}{\partial\langle\alpha\rangle}
=-2\langle\alpha\rangle\int\frac{d^3k}{(2\pi)^3}
\frac{1}{(k^2+\langle\bar{\tau}\rangle\langle\tau\rangle+\langle\sigma\rangle^2)^2}
-\left(\bar{v}_1 v_1-\bar{v}_2 v_2-\frac{1}{g}\right)=0,
\label{saddle3}\\
&&\frac{1}{N}\frac{\partial V}{\partial\langle\beta\rangle}
=\langle\bar{\beta}\rangle\int\frac{d^3k}{(2\pi)^3}
\frac{1}{(k^2+\langle\bar{\tau}\rangle\langle\tau\rangle+\langle\sigma\rangle^2)^2}
-\bar{v}_1 v_2=0,
\label{saddle4}\\
&&\frac{1}{N}\frac{\partial V}{\partial v_1}
=(\langle\bar{\tau}\rangle\langle\tau\rangle+\langle\sigma\rangle^2-\langle\alpha\rangle)
\bar{v}_1=0,
\label{saddle5}\\
&&\frac{1}{N}\frac{\partial V}{\partial v_2}
=(\langle\bar{\tau}\rangle\langle\tau\rangle+\langle\sigma\rangle^2+\langle\alpha\rangle)
\bar{v}_2=0.
\label{saddle6}
\eeqa
We look for the supersymmetric vacuum and have set
$-\langle\alpha\rangle^2+\langle\bar{\beta}\rangle\langle\beta\rangle=0$.
The solution of (\ref{saddle1}) through (\ref{saddle6}) is
\beq
\bar{v}_1 v_1=1/g,~~~v_2=0,
~~~\langle\tau\rangle=\langle\sigma\rangle=\langle\alpha\rangle=\langle\beta\rangle=0.
\label{solution}
\eeq
The saddle point conditions (\ref{saddle3}) and (\ref{saddle4}) contain the infrared divergences
when $\langle\tau\rangle=\langle\sigma\rangle=0$.
In deriving the solution (\ref{solution}) we have introduced the infrared cutoff.
Since the $\phi^\alpha_1$ fields have a vacuum expectation value,
$SU(N)$ symmetry of the action is broken.
The model has only the broken phase.
Performing the shift
\beq
\phi^N_1\to\phi^N_1+\sqrt{N}v_1
\label{phishift}
\eeq
in (\ref{fullaction}) and calculating the effective action,
we obain
\beqa
S_{eff}&=&N{\rm Tr}\ln\Delta_{B1}+N{\rm Tr}\ln\Delta_{B2}
-N{\rm Tr}\ln\Delta_{F1}-N{\rm Tr}\ln\Delta_{F2}
\nonumber\\
&&+N\bar{v}_1 v_1\int d^3x (A_\mu A_\mu+A_\mu\partial_\mu\Delta_{B1}^{-1}\partial_\nu A_\nu
+\bar{\tau}\tau+\sigma^2-\alpha\Delta_{B1}^{-1}\alpha-\beta\Delta_{B2}^{-1}\bar{\beta}
\nonumber\\
&&-\bar{c}\Delta_{F1}^{-1}c-\bar{e}\Delta_{F2}^{-1}e)
+\frac{N}{g}\int d^3x ~\alpha+\dots.
\label{effaction2}
\eeqa

The two-point functions of the auxiliary fields are obtained by evaluating functional derivatives
of the effective action (\ref{effaction2}).
All of these functions are finite because of cancellations between boson and fermion loops.
This is in accord with the renormalizabiliaty of the model.
The effective propagators of the auxiliary fields are given by
\beqa
&&D^A_{\mu\nu}(p)=\frac{1}{N}\frac{4}{\sqrt{p^2}+8\bar{v}_1 v_1}
\left(\delta_{\mu\nu}-\frac{p_\mu p_\nu}{p^2}\right),
\nonumber\\
&&D^\tau(p)=\frac{1}{N}\frac{8}{\sqrt{p^2}+8\bar{v}_1 v_1},
~~~D^\sigma(p)=\frac{1}{N}\frac{4}{\sqrt{p^2}+8\bar{v}_1 v_1},
\nonumber\\
&&D^\alpha(p)=-\frac{1}{N}\frac{4p^2}{\sqrt{p^2}+8\bar{v}_1 v_1},
~~~D^\beta(p)=-\frac{1}{N}\frac{8p^2}{\sqrt{p^2}+8\bar{v}_1 v_1},
\nonumber\\
&&D^c(p)=\frac{1}{N}\frac{8p\!\!\!/}{\sqrt{p^2}+8\bar{v}_1 v_1},
~~~D^e(p)=\frac{1}{N}\frac{8p\!\!\!/}{\sqrt{p^2}+8\bar{v}_1 v_1}.
\label{propagator}
\eeqa
We note that the mixing terms between $A_\mu$ and $\alpha$, $\sigma$ vanish
in the effective action (\ref{effaction2})
as those in the supersymmetric $CP^{N-1}$ sigma model in three dimensions \cite{Inami}.
In the present model the fields $\psi^\alpha_i$ remain massless,
so the term which involves $\epsilon_{\mu\nu\rho}$ is not induced in the two-point function of $A_\mu$.
Such a term is induced in the symmetric phase
of the supersymmetric $CP^{N-1}$ sigma model in three dimensions
where the fermion fields become massive \cite{Inami}.

\section{The $\beta$-Function and Finiteness}

The saddle point conditions have turned out to be free from UV divergences.
The coupling constant is not renormalized at leading order,
so the $\beta$-function of the model is given by
\beq
\beta(\tilde{g})=\tilde{g},
\label{betafunction}
\eeq
where $\tilde{g}$ is the dimensionless coupling constant.
The $\beta$-function has no fixed point and there is no phase transition.
This $\beta$-function differs from that of the $O(N)$, $CP^{N-1}$ sigma models
and their supersymmetric versions.
In these models only the coupling constant is renormalized at leading order,
and the $\beta$-function of these models is given by
$\beta(\tilde{g})=\tilde{g}(1-\tilde{g}/\tilde{g}_c)$,
where $\tilde{g}_c$ is the critical point.

We have confirmed that the superficially divergent two- and three-point functions
of the auxiliary fields are all finite at leading order. 
Therefore,
the model is finite to leading order in $1/N$.
This result is same as that of the two-dimensional model,
where the supergraph techniques are used \cite{Rocek}.

It is an important question
whether the finiteness of the model persists to higher orders in $1/N$.
This question can be invesigated to next-to-leading order in the same manner
as the supersymmetric $CP^{N-1}$ sigma model \cite{Inami}.

\section*{Acknowledgements}

This work is supported partially by the Grants in Aid of Ministry of Education,
Culture and Science
(Priority Area B "Supersymmetry and Unified Theory" and Basic Research C).
M. Y. is supported by a Research Assistant Fellowship of Chuo University.

\end{document}